# Does DeFi remove the need for trust?
# Evidence from a natural experiment in stablecoin lending

Kanis Saengchote*, Talis Putniņš and Krislert Samphantharak

*Chulalongkorn Business School*
*University of Technology Sydney*
*University of California San Diego*

This version: 13 July 2022

***ABSTRACT***

Decentralized Finance (DeFi) is built on a fundamentally different paradigm: rather than having to trust individuals and institutions, participants in DeFi potentially only have to trust computer code that is enforced by a decentralized network of computers. We examine a natural experiment that exogenously stress tests this alternative paradigm by revealing the identities of individuals associated with a DeFi protocol, including a convicted criminal. We find that, in practice, DeFi does not (yet) fully remove the need for trust in individuals. Our findings suggest that that because smart contracts are incomplete, they are subject to run risk (Allen and Gale, 2004) and personal character and trust of individuals are still relevant in this alternative financial system.

---

* Corresponding author. Chulalongkorn Business School, Chulalongkorn University, Phayathai Road, Pathumwan, Bangkok 10330, Thailand. (email: kanis@cbs.chula.ac.th).

0

# 1. Introduction

The financial system relies heavily on trust. Participants must trust their counterparties (individuals or institutions) to uphold their side of financial contracts, and as a fallback, participants trust that the legal system will enforce these contracts.[1] This necessity for trust is why financial systems tend to be larger and more developed in countries that have more social capital, higher trust, and stronger legal systems (e.g., Guiso et al., 2004).

In contrast, the rapidly growing Decentralized Finance (DeFi) system is built on a different paradigm: individuals' financial obligations are defined in "smart contracts" (computer code stored on the blockchain) that are enforced by a decentralized network of computers. The important feature here is that participants do *not* have to trust other individuals, and individual identities should *not* matter. Participants instead must trust computer code, and trust that the blockchain's decentralized computer network will continue to enforce the code. This difference has enormous implications because by removing the need for trust between individuals and reliance on a legal system, DeFi potentially enables financing and trade where it was previously not possible. Proponents of DeFi describe this property as "code is law", but does this property hold up in practice? Is it possible to have a financial system without the need for trust between individuals?

This paper sheds the first light on these issues by analyzing a recent natural experiment in which trust between individuals participating in a DeFi platform was called into question by an exogenous event. Specifically, we examine a bank-like DeFi protocol known as Abracadabra.Money. Abracadabra, like most DeFi protocols, works via a collection of smart contracts. These contracts accept the deposits of users, issue and record loans or depository receipts, monitor collateral value, undertake liquidations algorithmically if needed, and generally manage (in an automated fashion) the mechanics of the protocol. Effectively, the "shadow banking" services that are provided by Abracadabra are automated and hard-coded into its smart contracts, which are fully transparent in that every line of code can be viewed on the blockchain at the corresponding contract addresses. Further, the Abracadabra base smart contract has been audited by a security firm.[2] Therefore, the protocol, much like the DeFi ecosystem, is designed to operate such that a user need not trust any of the other users, nor does a user need to trust the founders or associated parties because the mechanics of the protocol, and mutual obligations are spelled out in transparent computer code. The identities of the individuals behind the protocol creation should not be relevant if "code is law."

---

[1] Contracts in financial systems are upheld by trust and enforced in accordance with the legal system (e.g., La Porta et al., 1997). Less trusting individuals are less likely to participate in financial markets (e.g., Guiso et al., 2008). Trust is also crucial in delegated investment management, which is a major component of most wealth/savings management (Gennaioli et al., 2015).

However, as we show in this paper, the identities were regarded as relevant by participants. When an individual associated with the Abracadabra protocol was identified as being a convicted criminal, that information triggered a collapse of trust and subsequent "run" on the protocol, illustrating this paper's central point that code has not (yet) replaced the need for trust.

We attribute this finding to the incompleteness of contracts, their inherent inability to cover all possible scenarios when providing complex services, and therefore elements of human discretion that are still present in DeFi systems. If markets and contracts were complete, there would be no need for "firms" (Williamson, 1979), computer codes would be sufficient to enforce all contractual relationships, and interpersonal trust would not matter.[3] We show that DeFi protocols ultimately issue incomplete contracts in the spirit of Allen and Gale (2004). Scope for human discretion, as a response to contract incompleteness and technical limitations, prevents DeFi protocols from fully disposing of the need for trust.

## 2. The natural experiment

Abracadabra was co-founded by Daniele Sestagalli, a serial tech entrepreneur, along with anonymous members identified only by pseudonyms (such as "Merlin"). Figure 1 provides a high-level overview of how the protocol works. Abracadabra allows participants to leverage their positions to amplify DeFi yields: (i) users deposit crypto assets (tokens) from yield-generating protocols (also known as interest-bearing tokens) as collateral, (ii) borrow the protocol's own stablecoin, known as "Magic Internet Money" (MIM; Panel A), and then (iii) MIM can be converted to other stablecoins via decentralized exchanges, which can then be cycled back into yield-generating protocols and the cycle repeated to generate leveraged investment. Participants are required to overcollateralize their positions to reduce credit risk and Abracadabra can automate the recursive interactions that leverage up yield (Panel B).[4] Panel C displays the balance sheets of related parties.[5]

---

[2] Most of Abracadabra's contracts are based on Sushi protocol's BentoBox contract which has been audited by Certora, a smart contract security firm. https://docs.abracadabra.money/our-ecosystem/our-contracts

[3] In traditional finance, incompleteness can be remedied by intermediaries, e.g., ex-post renegotiation (Hart and Moore, 1988), facilitated by social norms or legal frameworks, but this is not possible in in DeFi as contingencies need to be specified ex-ante as codes are often immutable and legal enforcement is limited.

*Figure 1 about here*

Abracadabra accepts a range of interest-bearing tokens from other protocols including Yearn, Terra USD (UST) stablecoin, and notably (for reasons that will become clear soon) Wonderland DAO, which is a yield-generating protocol also co-founded by Sestagalli. Each interest-bearing token has its own pool in Abracadabra. Most Abracadabra pools and contracts are deployed on the Ethereum blockchain but because the Wonderland protocol is built on the Avalanche blockchain, Abracadabra also deploys contracts on Avalanche for compatibility.[6]

Figure 2 shows MIM loans across both blockchains; by the end of December 2021, there are more than $3.1 billion of loans with almost $5.3 billion of collaterals.

*Figure 2 about here*

At 6:47am UTC on January 27, 2022, Twitter user @zachxbt revealed that Wonderland's anonymous treasury manager identified as "0xSifu" is in fact co-founder of QuadrigaCX, a Canadian exchange charged by authorities as Ponzi scheme and has criminal record on a separate count.[7] One minute before @zachxbt's tweet, Sestagalli tweeted that he has been "aware of this and decided that the past of an individual doesn't determine their future".[8] However, the community was not convinced: as treasurer, 0xSifu had access to Wonderland's funds which on January 11, 2022, totaled more than $1 billion of non-Wonderland tokens.[9]

---

[4] Recursive borrowing is a common use of leverage in DeFi. Saengchote (2021) studies Compound, one of the early DeFi lending protocols, and finds evidence of recursive borrowing. While recursive borrowing is often done manually, Abracadabra automates the process, allowing participants to borrow multiple times their initial capital. The on-chain token exchange can be done via swap pools intermediated via a "bonding curve" (Lehar and Parlour, 2022).
[5] From the perspective of endogenous money (Tobin,1970; Freeman and Kydland, 2000), MIM can be viewed as dollar-denominated transferable deposits created as credit via collateralized debt positions (CDP), making Abracadabra a crypto-based "shadow bank" (Li and Mayer, 2021).
[6] Avalanche is an EVM-compatible blockchain (Ethereum Virtual Machine), whose programming works similarly to Ethereum. It is possible to bridge tokens across other blockchains to allow interoperability, which is how TerraUSD (UST) stablecoin on the Terra blockchain interacts with Abracadabra on the Ethereum blockchain, as described in the Appendix.
[7] The full Twitter thread can be read at https://twitter.com/zachxbt/status/1486591682728673282, and news of QuadrigaCX in 2020 can be read at https://www.reuters.com/article/us-crypto-currencies-quadriga-idUSKBN23I3AF.
[8] The full Twitter thread can be read at https://twitter.com/danielesesta/status/1486591436233404421.

Anxious participants fled the Wonderland protocol, causing the price of wMEMO tokens (the interest-bearing token of Wonderland) to drop, leading to wMEMO-MIM liquidation cascades and withdrawals. From the peak in late December to end of January 2022, wMEMO-MIM loans declined by 97%.

Wonderland was at the time linked to Abracadabra only as collateral in wMEMO-MIM pools. As noted above, different collateral tokens were stored in separate pools in Abracadabra and governed by separate smart contracts designed to operate independently. Therefore, 0xSifu's role as treasurer of Wonderland was not directly relevant to Abracadabra. Thus, this event is a natural experiment that exogenously shocks DeFi participants' trust in Wonderland and Abracadabra through a shared co-founder.

By the end of January, MIM loans on Ethereum had declined by 48.1% from the peak. The magnitude is significant and could be considered a contagion, similar to runs on money market funds unrelated to Lehmann Brothers in 2008 documented by Akay et al. (2015).

## 3. Empirical analysis of the "run" following the shock to trust

We analyze contracts deployed on Ethereum and Avalanche and obtain hourly data on pool states such as outstanding MIM loans and collateral. This allows computation of loan-to-value ratios (LTV), which are report in Figure A2 in the Appendix. We obtain hourly token prices from the CoinGecko API.

Table 1 reports the pool snapshots at the end of December 2021, January 2022, all-time high, and change from all-time high to end of January. Table 1 and Figure 2 show that the magnitudes of loan repayments (withdrawals from Abracadabra) following the exogenous shock to trust are significant, with stablecoin pools declining by more than 50% and non-stablecoin pools more than 40%.

---

[9] A tweet on January 11, 2021, references to the non-native tokens worth over $1.05 billion held in the treasury. https://twitter.com/wonderland_fi/status/1480880342181494785

*Table 1 about here*

We use the Bai and Perron (2003) structural break algorithm to formally timestamp the spillover between protocols as information is publicly revealed. We apply the algorithm to detect structural breaks in a period of 96 hours (4 days) before and after the tweet time at 6:47am on January 27, 2022. The time axis is relabeled so that 0 represents the tweet hour, and negative values are hours before the tweet. The relatively short windows and high frequency isolate the effects of the exogenous shock from other events. We estimate the Bai and Perron (2003) algorithm using the following multiple regression model with $m$ breakpoints (hence $m + 1$ regimes):

$$y_t = x_t'\beta + z_t'\delta_j + u_t \qquad (1)$$

for $t = T_{j-1} + 1, \ldots, T_j$ and $j = 1, \ldots, m + 1$. $y_t$ is the dependent variable, and $x_t$ and $z_t$ are vectors of covariates. In the original version, $x_t$ are covariates whose coefficients are invariant for the entire sample, while $z_t$ are covariates that can experience structural breaks. In our implementation, we allow all coefficients to vary and thus only specify $z_t$, not $x_t$. This is referred to as the pure structural change model. The algorithm applies ordinary least square (OLS) regression segment-by-segment and identifies breakpoints as linear combinations of segments with the lowest sums of squared residuals (SSR). To address potential sensitivity to input parameters, we estimate breakpoints using minimum segment size $h$ of 5% and 10% of the sample.

For loan pool sizes and wMEMO price, we regress log value ($y_t$) on a constant, lagged log Ether price (ETH) to control for market conditions, and $y_{t-1}$ to allow for an AR(1) process in Equation 2. For stablecoin prices and pool LTV, we regress only on a constant because a stablecoin is intended to track $1 and pool LTV is designed to be stable.

$$y_t = \alpha + \beta_1 \ln(ETH_{t-1}) + \beta_2 y_{t-1} + u_t \qquad (2)$$

*All-pools*

Table 2 reports the breakpoints estimated with $h$ of 5% and 10%. For all pools except one, they are the same for both. The only pool with a different breakpoint is the deprecated Wonderland's wMEMO pool, where the 5% result is at a later hour than the tweet. With 10%, the results are the same. The breakpoints are identified at -31 hours, suggesting that insiders may have acted before information became public. In the tweet, @zachxbt posted screenshots of the conversation with Sestagalli timestamped at 5:32am "today", but the date and time zone are not verifiable from the screenshots, so it is unclear whether this chat takes place an hour before the tweet or much earlier, but this result exemplifies the importance informational advantage that is often subjected to regulation in traditional finance.

For most Ethereum pools, the breakpoints are many hours after the tweet, with exceptions of yvYFI, cvxrenCRV, wsOHM, and ALCX which account for 2.1%. For UST – the main stablecoin accounting for over 41% of loans – the breakpoints are 24 and 27 hours.

*Table 2 about here*

*Wonderland pools*

The algorithm can identify several breakpoints, so we allow up to 3 to be identified with $h$ of 5%. We focus on Wonderland variables: wMEMO price, wMEMO price relative to ETH (to control for market conditions), MIM loans in the two pools, and joint LTV. Figure 3 plots the breakpoints with the corresponding time series data, where dotted lines are breakpoints and bold lines are tweet time. For wMEMO price and relative price, they are (-33, -24, -1) and (-43, -33, 0) respectively, while for MIM loans and LTV, (-33, -16, 7) and (-32, -1, 38). The common breakpoints near the -33 hour are consistent with the insider result earlier: wMEMO price falls before the tweet, the fall is uncorrelated to the market, and the magnitude of the fall is significant. Pool LTV increases as deleveraging occurs, likely to due to closures of highly levered positions.

*Figure 3 about here*

*Ethereum pools*

In Figure 4, we repeat the multiple-breakpoint estimation on Ethereum. Panel A and B report the non-stablecoin and non-stablecoin loans breakpoints at (2, 15, 24) and (12, 21, 31), and Panel C reports the UST stablecoin at (5, 14, 25). However, visual inspection of the stablecoin (Panel B) suggests that (0) could also be considered a breakpoint but not identified by the algorithm.

*Figure 4 about here*

*Overall*

The results indicate substantial and rapid deleveraging, equating to participants exiting the Abracadabra protocol, after the tweet. The timestamps of structural breaks establish that the erosion of trust spreads from Wonderland to Abracadabra – likely through shared management – even though they are not explicitly connected by computer code.

## 4. Conclusion

Trust could be defined as the subjective probability individuals attribute to the possibility of being cheated. While DeFi has the potential to remove the risk of being cheated though the use of smart contracts that define rights and obligations and are enforced by a decentralized network, our analysis of a natural experiment with an exogenous shock reveals that in practice that is not yet a reality. As our analysis shows, personal identities and what they reveal about individuals are relevant, suggesting market participants place trust in individuals and a loss of that trust can lead to DeFi protocol runs akin to bank runs. Virtually all jurisdictions have fitness and propriety rules (F&P) that bar those who are unfit for office from holding senior positions in banks (Blinco et al., 2020), so perhaps similar regulations might be necessary for DeFi protocols.

Our results illustrate a limitation of the concept that "code is law" when contracts are incomplete in the sense of Allen and Gale (2004). The inability to anticipate and provision for every possible state ex-ante makes contracts imperfect and leads to elements of human discretion being embedded in DeFi systems to deal with unanticipated states. These include manual backstops, "multisig" authorizations of actions in protocols, ability to modify or pause smart contracts, override them with governance votes, and so on.[10] This state of practice has led some observers to label DeFi as "DINO" (decentralization in name only). Further, it is not always easy to discern what manual, discretionary mechanisms exist within a protocol and anticipate how they could be misused to the detriment of protocol participants. These realities, we argue, is why DeFi is not yet free of the need for trust.

---

[10] A DeFi architect remarks about how "just because a project has a few governance protocols doesn't necessarily mean they are decentralized". https://cointelegraph.com/news/wonderland-s-treasury-saga-exposes-the-fragility-of-dao-projects-today

**Table 1: Abracadabra Loan Snapshots**

This table reports summary statistics of loans in all Ethereum-based and Avalanche-based pools (wMEMO). The statistics are extracted from on-chain data of the blockchains. Loans are denominated in million units of MIM (Magic Internet Money, a stablecoin that acts as units of loans from Abracadabra). The maximum loans between the pool's inception and end of January 2022 are reported. We also report loan amounts at the end of December 2021 and January 2022 for comparison. The change from peak is calculated as the change from maximum loans to the end of January 2022. We separate the tokens by whether they are non-stablecoins are non-stablecoins (pegged to USD) and report the summary statistics for the older (deprecated) versions of the pools separately. Some tokens on the Ethereum blockchain are bridged across from their native blockchains, such as wBTC (Bitcoin is native to the Bitcoin blockchain) and UST (native to the Terra blockchain). Details of how UST pools are deployed (as wrapped UST or wUST) on the Ethereum blockchain in collaboration are explained in Figure A1. Panel A reports the statistics for non-stablecoins on the Ethereum blockchain, Panel B for stablecoins on the Ethereum blockchain, Panel C for Wonderland DAO's tokens on the Avalanche blockchain, and Panel D reports the aggregate statistics across the two blockchains.

| | All-Time High | End of Dec 2021 | Share | End of Jan 2022 | Share | Change from High |
|---|---|---|---|---|---|---|
| **Panel A: Non-stablecoins in Ethereum-based pools** | | | | | | |
| yvstETH | 216.93 | 216.03 | 7.1% | 93.91 | 5.8% | -56.7% |
| yvWETH | 209.08 | 199.50 | 6.5% | 131.28 | 8.1% | -37.2% |
| yvYFI | 0.23 | 0.23 | 0.0% | 0.16 | 0.0% | -32.5% |
| FTT | 69.78 | 57.25 | 1.9% | 69.48 | 4.3% | -0.4% |
| FTM | 55.39 | 55.35 | 1.8% | 52.24 | 3.2% | -5.7% |
| cvxrenCRV | 50.71 | 50.41 | 1.6% | 8.57 | 0.5% | -83.1% |
| wsOHM | 12.95 | 12.71 | 0.4% | 0.05 | 0.0% | -99.6% |
| xSUSHI | 27.80 | 25.43 | 0.8% | 4.33 | 0.3% | -84.4% |
| SPELL | 7.57 | 4.38 | 0.1% | 2.66 | 0.2% | -64.9% |
| sSPELL | 58.09 | 42.96 | 1.4% | 27.52 | 1.7% | -52.6% |
| wETH | 50.37 | 12.18 | 0.4% | 21.74 | 1.3% | -56.8% |
| wBTC | 41.24 | | | 23.12 | 1.4% | -43.9% |
| ALCX | 2.86 | 1.69 | 0.1% | 1.05 | 0.1% | -63.4% |
| Sum | 740.48 | 678.12 | 22.2% | 436.08 | 27.0% | -41.1% |
| **Panel B: Stablecoins in Ethereum-based pools** | | | | | | |
| cvxtricrypto2* | 352.65 | 336.32 | 11.0% | 164.43 | 10.2% | -53.4% |
| cvx3pool - depr | 369.27 | 361.96 | 11.8% | 40.47 | 2.5% | -89.0% |
| cvx3pool | 303.05 | 301.18 | 9.8% | 57.88 | 3.6% | -80.9% |
| yvUSDC | 54.77 | 54.77 | 1.8% | 51.58 | 3.2% | -5.8% |
| yvUSDT | 32.21 | 31.92 | 1.0% | 31.34 | 1.9% | -2.7% |
| yvcrvIB | 29.17 | 29.12 | 1.0% | 29.16 | 1.8% | 0.0% |
| UST - depr | 156.42 | 154.47 | 5.0% | 67.09 | 4.2% | -57.1% |
| UST | 1,116.50 | 1,112.89 | 36.4% | 737.99 | 45.7% | -33.9% |
| Sum | 2,394.24 | 2,382.63 | 77.8% | 1,179.94 | 73.0% | -50.7% |
| **Panel C: Tokens in Avalanche-based pools** | | | | | | |
| wMEMO - depr | 4.90 | 4.84 | 4.6% | 0.02 | 0.5% | -99.7% |
| wMEMO | 111.77 | 101.23 | 95.4% | 3.09 | 99.5% | -97.2% |
| **Panel D: Totals by blockchain** | | | | | | |
| Total Ethereum pools | 3,134.72 | 3,060.74 | 100.0% | 1,616.02 | 100.0% | -48.4% |
| Total Avalanche pools | 116.20 | 106.07 | 100.0% | 3.11 | 100.0% | -97.3% |

**Table 2: Structural Breaks Identification**

This table reports the dates for the structural break for outstanding loans identified by the Bai and Perron (2003) algorithm. The sample period is 96 hours (4 days) before and after the tweet by @zachxbt at 6:47am UTC on January 27, 2022. We define time in hours (t) on either side of this tweet, where time bucket beginning 7am is as classified as t=0. The breakpoints estimated using 5% and 10% of the sample in each regression are reported separately. The breakpoints are identified where the sums of squared residuals (SSR) of the regression is the lowest. We report the global SSR and the SSR at the breakpoints for both minimum segment size $h$ of 5% and 10% of the sample.

|  | Break (5%) | Break (10%) | SSR (global) | SSR (5%) | SSR (10%) |
|---|---|---|---|---|---|
| Panel A: Non-stablecoins in Ethereum-based pools | | | | | |
| yvstETH | 15 | 15 | 0.75 | 0.05 | 0.05 |
| yvWETH | 13 | 13 | 0.01 | 0.01 | 0.01 |
| yvYFI | -24 | -24 | 0.00 | 0.00 | 0.00 |
| FTT | 15 | 15 | 0.00 | 0.00 | 0.00 |
| FTM | 23 | 23 | 0.00 | 0.00 | 0.00 |
| cvxrenCRV | -74 | -74 | n.a. | 0.31 | 0.31 |
| wsOHM | -69 | -69 | 0.17 | 0.08 | 0.08 |
| xSUSHI | 48 | 48 | 1.43 | 0.02 | 0.02 |
| SPELL | 8 | 8 | 0.38 | 0.10 | 0.10 |
| sSPELL | 8 | 8 | 0.03 | 0.02 | 0.02 |
| wETH | 14 | 14 | 0.20 | 0.10 | 0.10 |
| wBTC | 18 | 18 | 0.12 | 0.04 | 0.04 |
| ALCX | -34 | -34 | 0.00 | 0.00 | 0.00 |
| Non-stablecoins (ETH) | 15 | 15 | 0.02 | 0.01 | 0.01 |
| Panel B: Stablecoins in Ethereum-based pools | | | | | |
| cvxtricrypto2* | 15 | 15 | 0.04 | 0.03 | 0.03 |
| cvx3pool - depr | 30 | 30 | 0.43 | 0.25 | 0.25 |
| cvx3pool | 48 | 48 | 0.49 | 0.29 | 0.29 |
| yvUSDC | 15 | 15 | 0.00 | 0.00 | 0.00 |
| yvUSDT | 14 | 14 | 0.00 | 0.00 | 0.00 |
| yvcrvIB | 39 | 39 | 0.02 | 0.02 | 0.02 |
| UST - depr | 27 | 27 | 0.05 | 0.04 | 0.04 |
| UST | 25 | 25 | 0.05 | 0.04 | 0.04 |
| Stablecoins (ETH) | 14 | 14 | 0.03 | 0.02 | 0.02 |
| Panel C: Tokens in Avalanche-based pools | | | | | |
| wMEMO - depr | 9 | -31 | 1.32 | 0.52 | 0.79 |
| wMEMO | -31 | -31 | 1.58 | 0.79 | 0.79 |

**Figure 1: High Level Overview of Abracadabra**

This figure illustrates how tokens and loans can be recursively used in the Abracadabra protocol to increase leverage and earn higher yields on initial collateral. Panel A illustrates the basic interaction where a token representing initial equity (e.g., USDT stablecoin) is staked to a yield aggregator protocol (e.g., Yearn) and converted to an interest-bearing token (e.g., yvUSDT). The conversion from USDT to yvUSDT is like converting cash into a bank deposit, and yvUSDT can be thought of as depository receipt. The interest-bearing token is then used as collateral to borrow the Magic Internet Money (MIM) stablecoin, which must be repaid with interest to release the collateral. Panel B illustrates a recursive use where the borrowed MIM is swapped to USDT and then staked to yvUSDT, which can then be further used as collateral to borrow more MIM. This recursive action is automated by the Abracadabra protocol and participants can choose the level of leverage, which determines their yield and risk of liquidation. For the case of TerraUSD (UST), Abracadabra officially collaborates with Terra to facilitate the bridging of tokens from the Ethereum blockchain to the Terra blockchain, so the yield-earning protocol is on the Terra blockchain while Abracadabra and swap pools are on the Ethereum blockchain (see Appendix). For other pools, the protocol is built to operate on the same chain. Panel C presents the balance sheet view of a participant, the protocol, and Curve.fi MIM+3Crv decentralized swap pool that can be used to swap MIM for other stablecoins. Bold lines denote tokens that are held and visible in blockchain addresses, and dotted lines are conceptual obligations.

**Panel A: basic interaction**

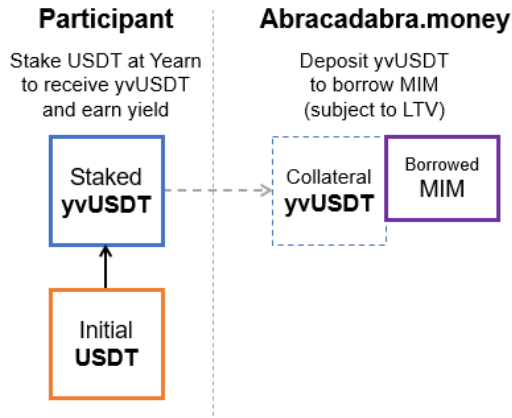

**Panel B: recursive interaction**

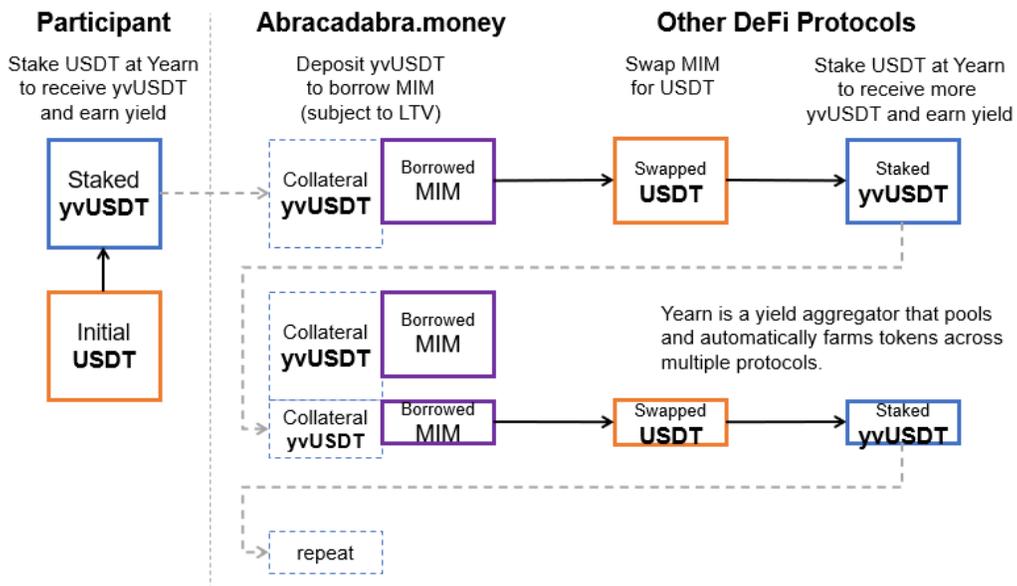

**Panel C: balance sheet view of participant and related protocols**

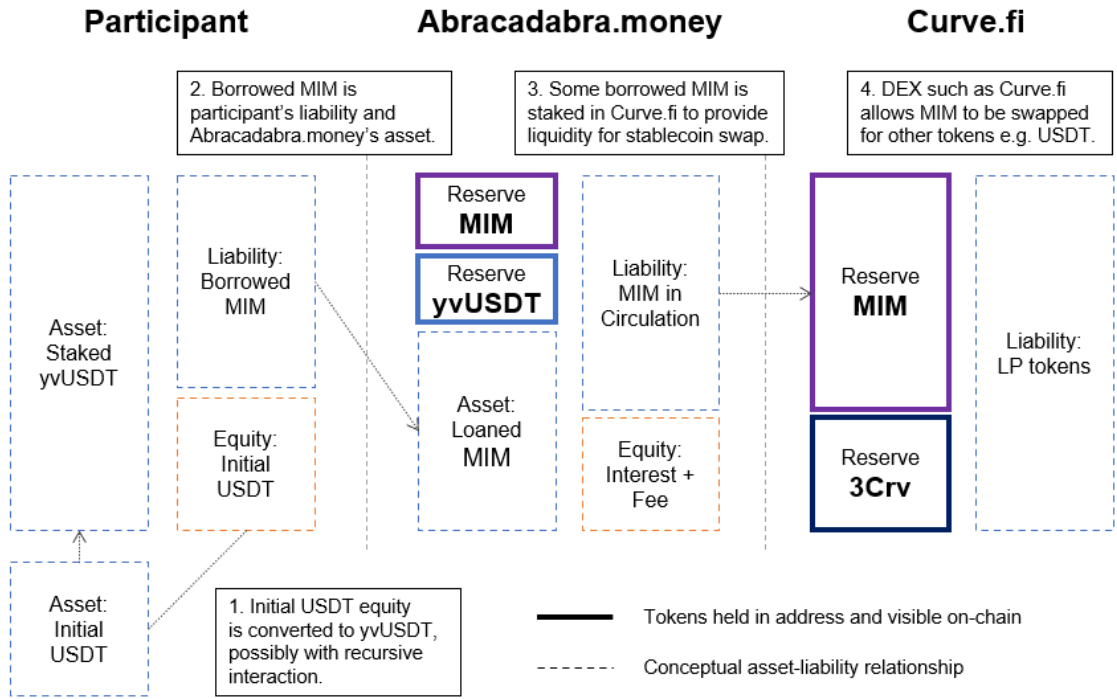

**Figure 2: Abracadabra Loans**

This figure plots the hourly outstanding loans in Abracadabra's Ethereum-based and Avalanche-based pools between June 2021 and January 2022. Aggregate loans are extracted from on-chain data of the blockchains. Loans are denominated in million units of MIM (Magic Internet Money, a stablecoin that acts as dollar-based promissory notes issued by Abracadabra). In Panel A, loans on the Ethereum blockchain are aggregated by loans backed by non-stablecoins and stablecoins, and TerraUSD (UST) is also reported as a subset as it accounts for the majority of stablecoin-backed loans. Panel B reports the loans on the Avalanche blockchain. The vertical lines in Panel A and B are the tweet time by @zachxbt at 6:47am UTC on January 27, 2022. The pool inception date for wMEMO is September 30, 2022, so the sample periods for Panel B and A are slightly different.

**Panel A: Ethereum pools**

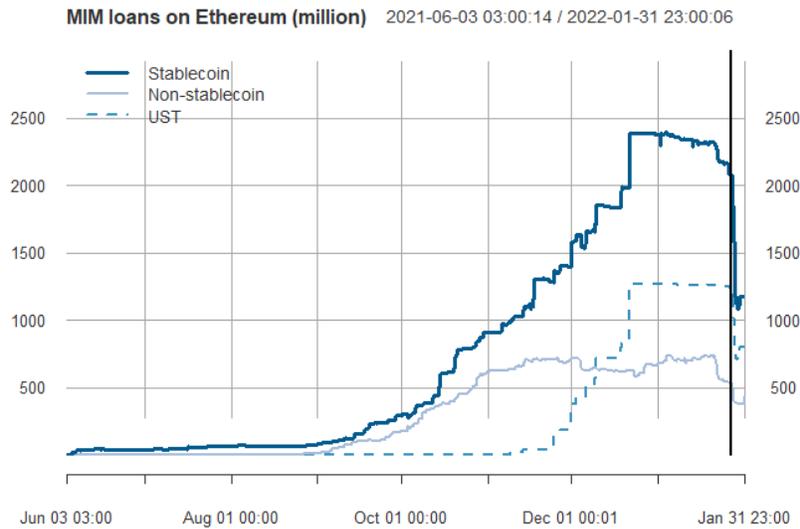

**Panel B: Avalanche pools (wMEMO)**

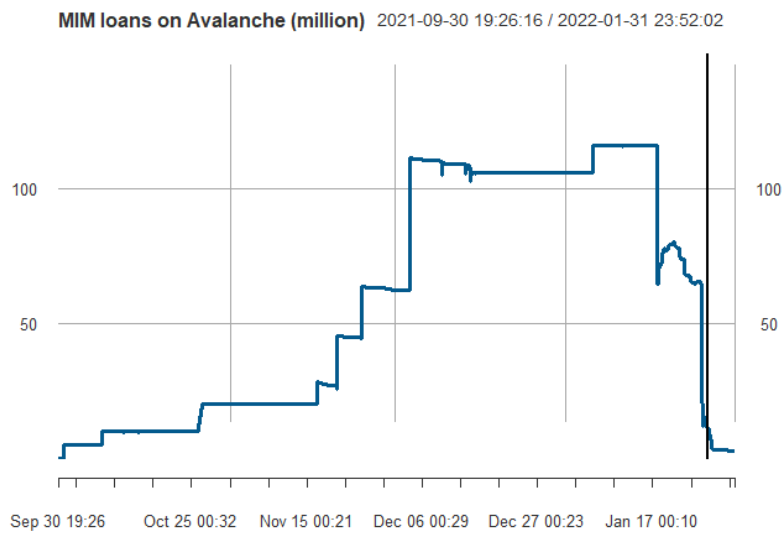

**Figure 3: Events on the Avalanche Blockchain**

This figure plots the hourly activities in Abracadabra's Avalanche-based pools 96 hours (4 days) before and after the tweet time by @zachxbt at 6:47am UTC on January 27, 2022. Panel A plots the price of the wMEMO, the governance token of Wonderland. Panel B plots the price of wMEMO relative to Ether (ETH) – the native coin of the Ethereum blockchain – to control for general market conditions. Panel C plots the aggregate amount of MIM loans backed by wMEMO as collateral across both pools, and Panel D plots the loan-to-value (LTV) ratio of the MIM loans relative to the wMEMO collateral in the pools. In all panels, the blue line is the t=0 baseline, and the dotted lined are the times identified by the Bai and Perron (2003) algorithm as structural breakpoints. We allow up to 3 breakpoints to be identified and use 5% of the sample for each regression.

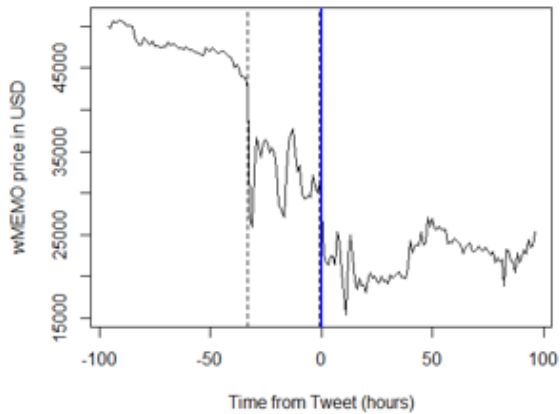
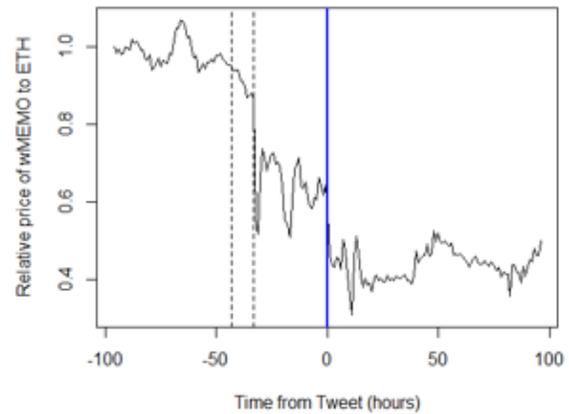
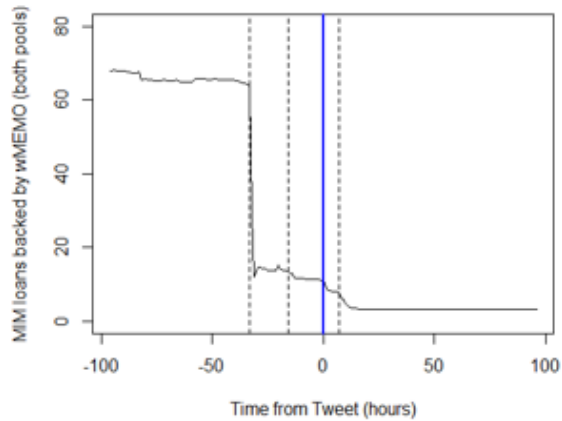
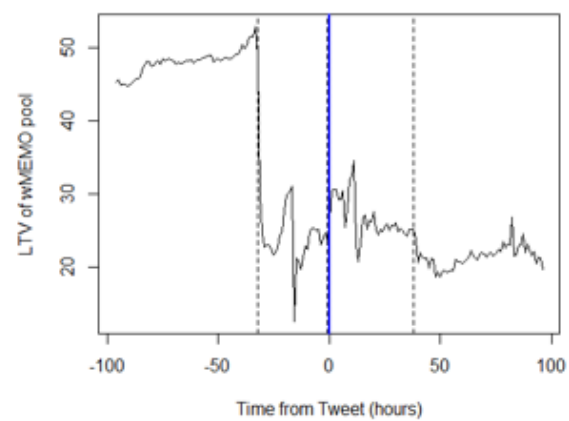

**Figure 4: Events on Ethereum Blockchain**

This figure plots the hourly activities in Abracadabra's Avalanche-based pools 96 hours (4 days) before and after the tweet time by @zachxbt at 6:47am UTC on January 27, 2022. Panel A and B plot the aggregate amount of MIM loans backed by non-stablecoin and stablecoin collateral respectively. Panel C plots the aggregate amount of MIM loans backed by TerraUSD (UST). In all panels, the blue line is the t=0 baseline, and the dotted lined are the times identified by the Bai and Perron (2003) algorithm as structural breakpoints. We allow up to 3 breakpoints to be identified and use 5% of the sample for each regression.

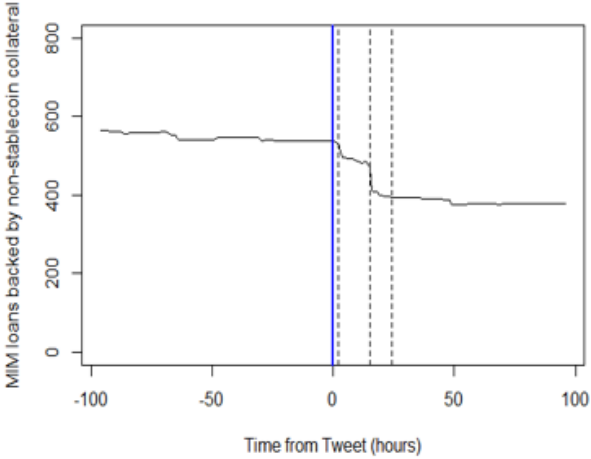
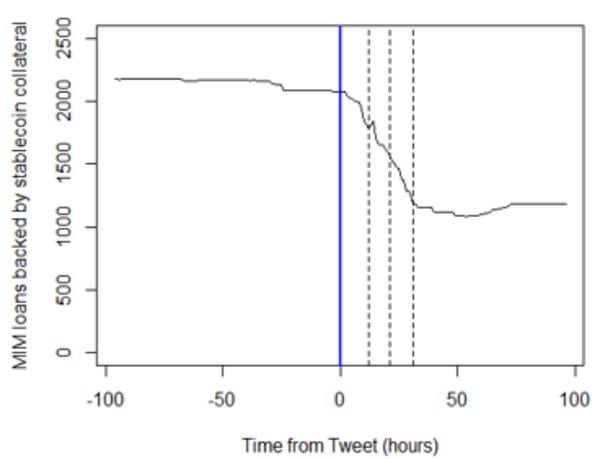
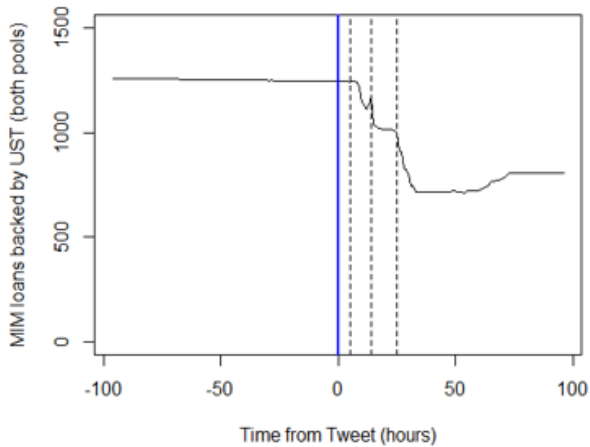

**Appendix: Further details of the Abracadabra protocol and structural break**

In October 2021, Terra and Abracadabra announced a cross-blockchain partnership to facilitate leveraged investment strategy. TerraUSD (UST) can be bridged between the Terra blockchain and the Ethereum blockchain so that Abracadabra's smart contract on Ethereum can lend MIM loans, which can then be converted to yield-generating tokens of Anchor savings and loan protocol on the Terra blockchain.[11] Details of how the collaboration works is described in Figure A1. The stability of stablecoin collateral-loan pair allows much higher leverage, as illustrated in Figure A2.

Soon after the launch, Twitter accounts begin offering ideas on how to leverage 20% returns on Anchor to more than 100%, which is extremely rare for stablecoins.[12] Popularity of UST-MIM loan skyrockets in December, reaching more than $1.2 billion by the end of December. By then, there are approximately $5 billion of UST deposited in Anchor, so at one point, Abracadabra accounts for almost a quarter of all deposits. LUNA's price – Terra blockchain's native coin that is used to mint UST – doubles in price between the launch of the UST pool and end of December. The relationship is plotted in Figure A2. LUNA suffers a 10% decline as participants flee Abracadabra in the wake of Wonderland's collapse but UST experiences only a small and momentary depeg as illustrated in Figure 5 Panel B.[13] The LUNA drop may be a result of the arbitrage mechanism that allows participants to redeem underpriced UST for $1 worth of newly minted LUNA. The increase in LUNA supply likely exerts negative price pressure on the native coin. This is the key feature of algorithmic stablecoins that make them vulnerable to runs, as studied by Saengchote and Samphantharak (2022).

We can use the structural break algorithm to analyze MIM and UST prices. Figure A4 shows that MIM and UST prices change at 15 and 17 hours respectively, close to the hours identified in Table 2 and Figure 4, but the magnitudes are small – less than 2 cents – similar to when Reserve Primary fund "breaks the buck" in 2008 (Akay et al., 2015). Compared to other algorithmic stablecoins such as Iron Finance's IRON and UST studied by Saengchote and

---

[11] Details of how the bridging process can be found in this Medium article. Wormhole is a multi-chain bridge that allows participants to bridge tokens across many blockchains. https://medium.com/terra-money/wormhole-v2-for-terra-the-ui-walkthrough-595ca6649ae8
[12] For example, see https://twitter.com/Route2FI/status/1458072000602202114.
[13] https://markets.businessinsider.com/news/currencies/luna-crypto-blockchain-scandal-money-fraud-markets-investors-retail-defi-2022-1

Samphantharak (2022), the MIM depeg is too small and short-lived to be called a stablecoin run. This highlights a crucial fact that not all stablecoins are equivalent and perform different roles within each protocol.

In Abracadabra, MIM is transferable deposit created as CDP credit, like MakerDAO's DAI studied by Kozhan and Viswanath-Natraj (2021), while UST is USD-pegged collateral supplied to the Abracadabra. MIM is required to repay loans and release locked collateral. When a borrower's CDP is insufficiently collateralized, third-party liquidators can buy MIM to repay on behalf and seize the collateral. But because MIM is liquid and transferable, it is also used as liquidity to swap stablecoins in Curve.fi. If participants observe this tweet and decide to exit Abracadabra, sudden sales of MIM would exert negative price pressure. The reduction in MIM price would make loan repayments "cheaper", so part of the deleveraging could be encouraged by this discount.

If the external collateral backing MIM is worth more than $1 per unit, MIM price should return to $1, unlike algorithmic stablecoins whose collateral are endogenous to the protocol. The loss of faith can eventually lead to algorithmic stablecoin runs (Saengchote and Samphantharak, 2022), making CDP stablecoins are more robust (Kozhan and Viswanath-Natra, 2021)

**Figure A1: Abracadabra and Terra collaboration**

This figure illustrates how TerraUSD (UST) stablecoins can be recursively used to increase leverage and earn higher yields on initial collateral. Rather than yield aggregator, Terra uses Anchor, a savings and loan protocol built on the Terra blockchain which allows participants to earn almost 20% on their stablecoin deposits. The interest-bearing token version of UST is aUST. Because Abracadabra is not built on the Terra blockchain, participants can start with a bridged (or "wrapped") version of UST on the Ethereum blockchain (wUST) and bridging it back to the Terra blockchain. The staked aUST is then bridged to the Ethereum blockchain (waUST) and used as collateral to borrow Magic Internet Money (MIM) stablecoin. The borrowed MIM can then be swapped for wUST, bridged to the Terra blockchain as UST and then staked as aUST, and so on. Because of the steps involved, the gas fee associated with this recursive lending can be high, but it can potentially enhance the rate of return from almost 20% to more than 100%, but participants will face greater liquidation risk for the same level of UST stablecoin price decline.

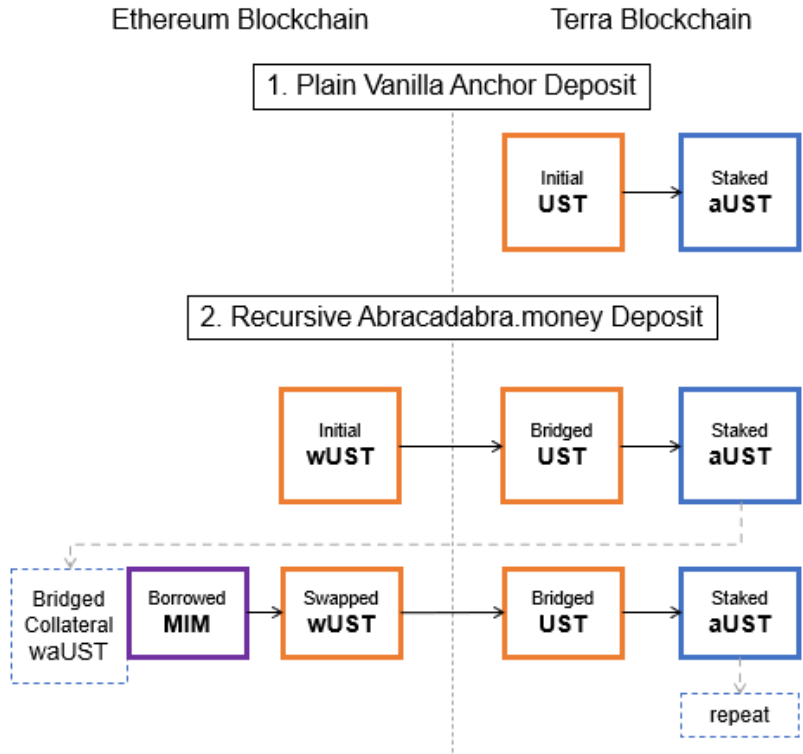

**Figure A2: Median pool LTV**

This figure plots the median loan-to-value (LTV) ratio of the MIM loans relative to the collateral in each pool from the pool inception date to then end of January 2022 at daily frequency. Aggregate loan and collateral values are extracted from on-chain data of the Ethereum and Avalanche blockchains. We separate the tokens by whether they are non-stablecoins (in blue), non-stablecoins (in orange) or on the Avalanche blockchain (in grey).

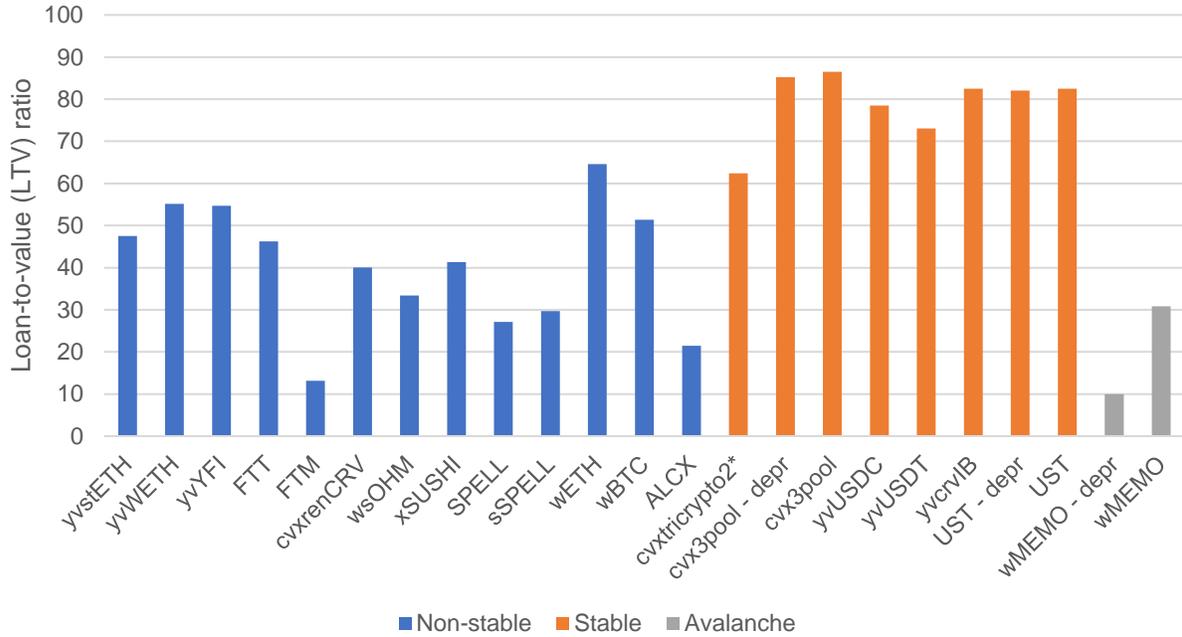

**Figure A3: Relationship between LUNA and UST-backed MIM loans**

This figure plots the hourly outstanding loans in Abracadabra's TerraUSD (UST) pools between November 2021 and January 2022. Aggregate loans are extracted from on-chain data of the Ethereum blockchain. UST loan is plotted in dotted line and reported on the righthand scale in million units of MIM (Magic Internet Money, a stablecoin that acts as dollar-based transferable deposit issued by Abracadabra). The price of LUNA, Terra blockchain's native coin which is used to mint TerraUSD (UST) stablecoin is plotted in solid blue line and reported on the righthand scale in US dollars. The price is obtained from CoinGecko.

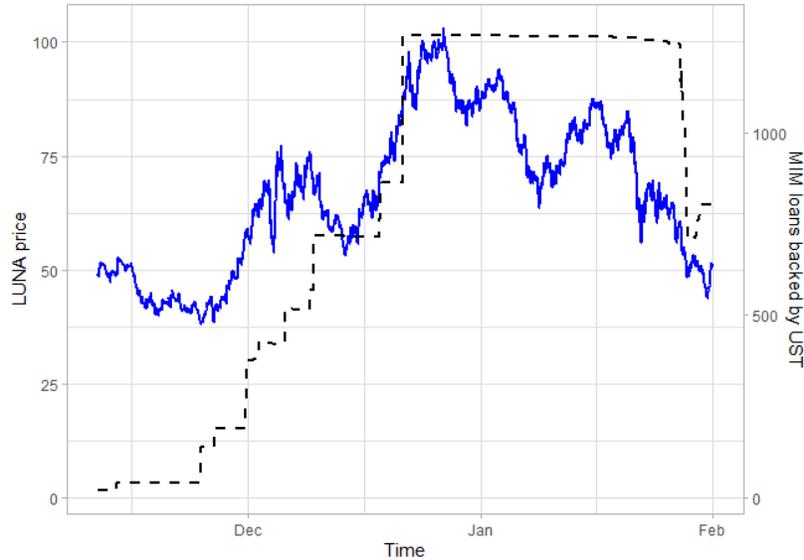

**Figure A4: Stablecoin Prices**

This figure plots the hourly prices of key stablecoins 96 hours (4 days) before and after the tweet time by @zachxbt at 6:47am UTC on January 27, 2022. Panel A plots the price of Magic Internet Money (MIM), a stablecoin which acts as dollar-based certificates of deposit issued by Abracadabra and panel B plots the price of TerraUSD (UST), an algorithmic stablecoin native to the Terra blockchain and bridged to the Ethereum blockchain. In both panels, the blue line is the t=0 baseline, and the dotted lined are the times identified by the Bai and Perron (2003) algorithm as structural breakpoints. We allow only a single breakpoint to be identified and use 5% of the sample for each regression.

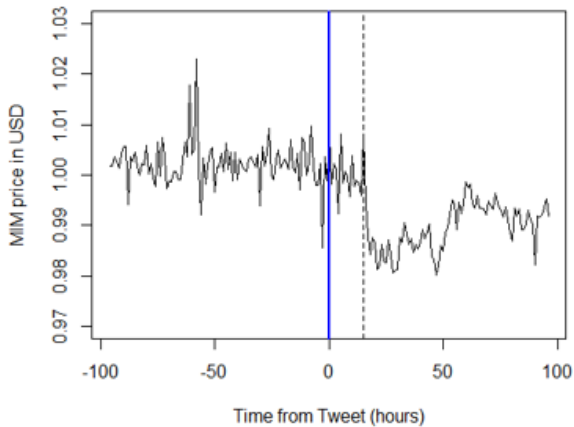
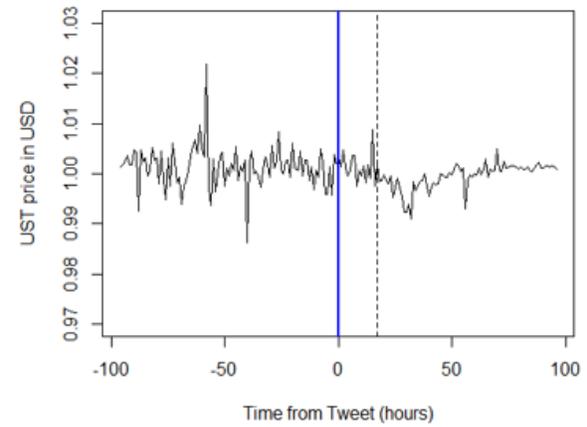